\documentclass{ifacconf}

\usepackage{graphicx}      
\usepackage{natbib}        

\usepackage{amssymb}
\usepackage{xurl}
\usepackage{amsmath}
\usepackage{amssymb}
\usepackage{amsfonts}
\usepackage{mathtools}
\usepackage{dsfont}
\usepackage{subfigure}
\usepackage{amsbsy}
\usepackage{mathrsfs}
\usepackage{lipsum}
\usepackage{color, soul}
\usepackage{lineno}
\usepackage{comment}
\usepackage[inkscapelatex=false]{svg}

\usepackage{tikz}
\usetikzlibrary{shapes, arrows, positioning,plotmarks,patterns}
    
\usepackage{pgfplots}
\pgfplotsset{
legend style={fill opacity=0.7,  draw opacity=1, text opacity=1, draw=white!15!black, legend cell align=left, align=left},
width=6cm, 
height=6cm,
yminorticks=false,
xminorticks=false,
title style={font=\small},
tick style={color=black},
tick label style={font=\small},
grid style={line width=.1pt, draw=gray!20},
major grid style={line width=.1pt,draw=gray!20},
}
\usepgfplotslibrary{colorbrewer}
\usepgfplotslibrary{groupplots}
\usepgfplotslibrary{fillbetween}
\usetikzlibrary{patterns}
\pgfplotsset{
compat=1.11,
every tick label/.append style={font=\footnotesize},
legend image code/.code={
\draw[mark repeat=2,mark phase=2]
plot coordinates {
(0cm,0cm)
(0.15cm,0cm)        
(0.3cm,0cm)         
};%
}
}

\definecolor{aoenglish}{rgb}{0.0, 0.5, 0.0}
\definecolor{darkblue}{rgb}{0.0, 0.0, 0.55}
\definecolor{darkmagenta}{rgb}{0.55, 0.0, 0.55}
\definecolor{electricviolet}{rgb}{0.56, 0.0, 1.0}
\definecolor{electricyellow}{rgb}{1.0, 1.0, 0.0}
\definecolor{forestgreen}{rgb}{0.13, 0.55, 0.13}
\definecolor{fuchsia}{rgb}{1.0, 0.0, 1.0}
\definecolor{gamboge}{rgb}{0.89, 0.61, 0.06}
\definecolor{goldenpoppy}{rgb}{0.99, 0.76, 0.0}
\definecolor{indigo}{rgb}{0.29, 0.0, 0.51}
\definecolor{internationalorange}{rgb}{1.0, 0.31, 0.0}
\definecolor{lava}{rgb}{0.81, 0.06, 0.13}
\definecolor{selectiveyellow}{rgb}{1.0, 0.73, 0.0}
\definecolor{turquoiseblue}{rgb}{0.0, 1.0, 0.94}
\definecolor{turquoise}{rgb}{0.19, 0.84, 0.78}
\definecolor{gold}{RGB}{255,215,0}
\definecolor{ferngreen}{RGB}{0,140,69}
\definecolor{brightwhite}{RGB}{244,245,240}
\definecolor{flamescarlet}{RGB}{205,33,42}
\definecolor{mygray}{RGB}{200,200,200}
\definecolor{darkorange}{RGB}{255,100,0}
\definecolor{darkgreen}{RGB}{0,102,0}
\definecolor{darkred}{RGB}{139,0,0}
\definecolor{darkblue}{RGB}{0,0,153}
\definecolor{greenVale}{RGB}{0,102,0}
\definecolor{darkviolet}{RGB}{153,50,204}

\definecolor{color0}{HTML}{00429D}
\definecolor{color1}{HTML}{844D99}
\definecolor{color2}{HTML}{C3608E}
\definecolor{color3}{HTML}{EF8078}
\definecolor{color14}{HTML}{915a8f}
\definecolor{color34}{HTML}{d27f76}
\definecolor{color4}{HTML}{FFB047}

\DeclareMathOperator*{\argmin}{arg\,min}    

\usepackage{glossaries}

\newacronym{bss}{BSS}{Bike Sharing System}
\newacronym{sr}{SR}{Static Rebalancing}
\newacronym{ctmc}{CTMC}{Continuous-Time Markov Chain}
\newacronym{dtmc}{DTMC}{Discrete-Time Markov Chain}
\newacronym{mdp}{MDP}{Markov Decision Process}
\newacronym{drl}{DRL}{Deep Reinforcement Learning}
\newacronym{mit}{MIT}{Massachusetts Institute of Technology}

\protected\edef\ell{\noexpand\ensuremath{{\mathchar\the\ell}}}
\begin{document}
\begin{frontmatter}

\title{Fully Dynamic Rebalancing in Dockless Bike-Sharing Systems via Deep Reinforcement Learning\thanksref{footnoteinfo}} 

\thanks[footnoteinfo]{This work was partially carried out within the Italian National Center for Sustainable Mobility (MOST) and received funding from NextGenerationEU (Italian NRRP – CN00000023 - D.D. 1033 17/06/2022 - CUP C93C22002750006).}

\author[First]{Edoardo Scarpel} 
\author[First]{Alberto Pettena}
\author[First]{Matteo Cederle}
\author[First]{Federico Chiariotti} 
\author[First]{Marco Fabris} 
\author[First]{Gian Antonio Susto} 

\address[First]{University of Padua, via Gradenigo 6/B, 35131, Padua, Italy (e-mail: edoardo.scarpel@phd.unipd.it, alberto.pettena@studenti.unipd.it, matteo.cederle@phd.unipd.it, federico.chiariotti@unipd.it, marco.fabris.1@unipd.it, gianantonio.susto@unipd.it).}


\begin{abstract}    
This paper proposes a fully dynamic \gls{drl} method for rebalancing dockless bike-sharing systems, overcoming the limitations of periodic, system-wide interventions. We model the service through a graph-based simulator and cast rebalancing as a Markov decision process. A \gls{drl} agent routes a single truck in real time, executing localized pick-up, drop-off, and charging actions guided by spatiotemporal criticality scores. Experiments on real-world data show significant reductions in availability failures with a minimal fleet size, while limiting spatial inequality and mobility deserts. Our approach demonstrates the value of learning-based rebalancing for efficient and reliable shared micromobility.



\end{abstract}


\end{frontmatter}

\glsresetall


\section{Introduction}

Smart cities are complex systems in which mobility is intricately interwoven with economic activity and social interactions. Rapid urbanization has intensified these dynamics: over half of the world's population now lives in urban areas, a proportion expected to rise to $68\%$ by 2050 \citep{UN_WorldUrbanizationProspect_2018}. This growth exacerbates congestion, pollution, and iniquities in accessibility \citep{cervero2013transport}. 

Together with traffic flow monitoring~\citep{Fabris2025sensorselection} and visual sensor grids for civic surveillance~\citep{varotto2022visual}, shared mobility services such as \glspl{bss} play a critical role in sustainable urban mobility solutions. Indeed, these services complement public mass transit with a more capillary network~\citep{sipe2023firstlastmile,yang2023bikemetro}, increasing freedom of movement within the city and reducing car dependence~\citep{li2020docklessbikes} and the consequent pollution and traffic.

The over 3{,}000 \glspl{bss} throughout the world\footnote{\url{https://bikesharingworldmap.com/}} promote healthier travel habits and integrate with public transport \citep{midgley2009role, pucher2011bicycling}. However, maintaining adequate bike availability represents a major hurdle for widespread adoption: heterogeneous travel patterns lead to spatial and temporal imbalances, reducing service availability and user satisfaction \citep{raviv2013static}. Rebalancing operations are therefore essential and can be \textit{static}, following pre-planned vehicle routes \citep{chemla2013bike, o2015data}, or \textit{dynamic}, adapting in real time to evolving demand patterns \citep{schuijbroek2017inventory, chiariotti2020bike}.

Efficiently redistributing bikes in large-scale systems remains a challenging problem. Traditional optimization and heuristic methods often struggle to cope with stochastic, time-varying demand and complex operational constraints. To address these limitations, recent research has explored learning-based approaches capable of adapting to evolving system dynamics and improving rebalancing performance. \gls{drl} provides a natural framework for designing dynamic policies that anticipate and respond to demand fluctuations. Early \gls{drl} applications to \gls{bss} rebalancing have shown promising results, exploiting temporal and spatiotemporal patterns to optimize operations \citep{pan2019deep, yin2023deep, pan2023novel, liang2024reinforcement}, though they often simplify system constraints. Building on this perspective, our previous work \citep{cederle2025fairness} showed that learning-based rebalancing can also incorporate fairness considerations, providing a more equitable service by avoiding discrimination against underprivileged neighborhoods while maintaining a high operational efficiency. 

Nevertheless, rebalancing remains a task that is undertaken periodically over the whole network, often neglecting the time and effort necessary to reach multiple stations over a short period. To the best of our knowledge, the possibility of a \emph{fully dynamic} rebalancing scheme, in which a smaller fleet of trucks, or even a single truck, can perform pointwise interventions and place bikes where they are needed without major rebalancing operations, has not been investigated so far. Such a fully dynamic rebalancing would be able to reduce the impact of the trucks on traffic, as well as improving service availability and reducing costs.

In this work, we present such a fully dynamic scheme, using \gls{drl} to route the truck and control the placement and charging of shared electric bikes\footnote{Although we focus on e-bikes, the framework naturally extends to non-electric bike-sharing systems.} throughout the day. We design the \gls{mdp} and reward function by dividing areas into critical, stable, or in surplus, leading the truck to place bikes where they are needed even with a small bike fleet. Our results show that \gls{drl} can rebalance bikes effectively and fairly even with a comparatively small bike fleet, maintaining a high availability throughout the service area without mobility deserts or other iniquities.

The remainder of the paper is organized as follows. Section~\ref{sec:problem_formulation} presents the system model, while Section~\ref{sec:approach} details the proposed \gls{drl} approach, the design of the \gls{mdp}, and the neural network architecture. Section~\ref{sec:results} reports our simulation results and Section~\ref{sec:conclusions} concludes the paper.


\section{Problem formulation}\label{sec:problem_formulation}

This section establishes the mathematical framework for the bike-sharing rebalancing problem. We begin by introducing the graph-based demand model (\S\ref{sec:bs_graph}), notation for vehicle states and dynamics, and the fundamental operations available to the rebalancing truck (\S\ref{sec:reb_model}). We then formulate the dynamic rebalancing problem as a sequential decision process under stochastic demand (\S\ref{sec:fully_dynamic_rebalancing}).

\subsection{Bike sharing system demand model}
\label{sec:bs_graph}

We model a target rebalancing area of a wider \gls{bss} as a connected, directed, and weighted graph 
$\mathcal{G} = (\mathcal{V}, \mathcal{E})$ \citep{harary2018graph}, 
where the vertex set $\mathcal{V}$ represents locations at which bikes may be parked or requested. In this \textit{dockless} configuration, each node $n \in \mathcal{V}$, with $|\mathcal{V}|=N$, represents a geographical area, with a time-varying bike occupancy $S_n(t)$ at time $t \in \mathbb{R}^+$, and the global occupancy state is $\mathbf{o}(t) = \left(o_0(t),o_1(t), o_2(t), \dots, o_N(t)\right)$. Node $0$ is a fictitious node modeling trips to and from \gls{bss} locations outside the rebalancing area. Each location has a capacity $S_n^{\max} \in \mathbb{N} \cup \{\infty\}$, which is unbounded in the dockless case. 
The edge set $\mathcal{E} \subseteq \mathcal{V} \times \mathcal{V}$ contains directed edges $(n,m)$, representing travel routes from node $n$ to node $m$.
Pairwise distances $\ell_{nm}$ between adjacent nodes are collected in matrix $L \in \mathbb{R}^{N \times N}$. As $\mathcal{G}$ is directed, $L$ is generally asymmetric. 

We divide each day into $T$ discrete time slots $\tau \in \{1,2,\dots,T\}$, each corresponding to a real-time interval $\mathcal{I}_{\tau} = [t_\tau^{\text{start}}, t_\tau^{\text{end}})$, enabling a dynamic representation of mobility patterns across the day. We then model the \gls{bss} demand of trips over edge $(n,m)$ in time slot $\tau$ as a Markov-modulated Poisson process with rate $\lambda_{nm}(\tau)$, which can be computed empirically.  The expected inflow-outflow rates at each node are then
\begin{equation*} 
    \Lambda_n^{\text{in}}(\tau) = \sum_{\mathclap{m:(m,n)\in\mathcal{E}}} \lambda_{mn}(\tau), 
    \quad
    \Lambda_n^{\text{out}}(\tau) = \sum_{\mathclap{m:(n,m)\in\mathcal{E}}} \lambda_{nm}(\tau).
\end{equation*}
 If demand for a trip is generated in location $n$, it may be redirected to another close-by location within a maximum acceptable distance $d_{\text{walk}}$ if there are no available bikes.

Each e-bike $b \in \mathcal{B}$ has position $p_b(t) \in \mathcal{V}$ and residual battery charge $c_b(t) \in [0, C_{\max}^b]$. If its charge is sufficient to travel from node $n$ to node $m$, the battery decreases linearly with distance:
\begin{equation*}
    c_b(t+\varepsilon) = c_b(t) - \varepsilon \cdot v, 
    \qquad \varepsilon \in \left[0, \ell_{nm}v^{-1}\right],
    \label{eq:bike_discharge_eq}
\end{equation*}
where $v$ is the mean travel velocity, drawn from a truncated Gaussian distribution with mean $\mu_v$, variance $\sigma_v^2$, and lower/upper bounds $v_{\min}$, $v_{\max}$:
\begin{equation*}
    f(v\mid\mu_v, \sigma_v, v_{\min}, v_{\max}) =
    \frac{1}{Z \sigma_v} \varphi\!\left(\frac{v - \mu_v}{\sigma_v}\right),
    \label{eq:truncated_gaussian_eq}
\end{equation*}
where $z_{\min} = (v_{\min} - \mu_v)/\sigma_v$ and $z_{\max} = (v_{\max} - \mu_v)/\sigma_v$ are the standardized bounds, $Z = \Phi(z_{\max}) - \Phi(z_{\min})$ ensures normalization, and $\varphi(\cdot), \Phi(\cdot)$ are the standard normal PDF and CDF, respectively.

The number of available bicycles at node $n$ at time $t$, denoted as $o_n(t)$, is decomposed into \textit{charged} and \textit{depleted} subsets as $o_n(t) = o_n^{c}(t) + o_n^{d}(t)$. A bike $b$ is considered active if its charge $c_b(t) > 0.2C_{\max}^b$ and depleted otherwise, and depleted e-bikes are considered unusable until they are recharged. The network state $\mathbf{o}(t)$ is then rewritten and augmented as follows: 
\begin{equation*}
    \mathbf{o}^*(t) = (o_0^c(t),o_0^d(t), \ldots, o_N^c(t), o_N^d(t)).
\end{equation*}
The number of bikes \emph{en route} from node $n$ to node $m$ at time $t$ is modeled as $r_{nm}(t)$, and the values for all edges are collected in vector $\mathbf{r}(t)$. The full state vector is then
\begin{equation*}
    \mathbf{x}(t) = (\mathbf{o}^*(t), \mathbf{r}(t)).
\end{equation*}
In order to model redirected trips, we define the closest available node as
\begin{equation*}
g_n(t)=\argmin_{g\in\mathcal{V}:o_g^c(t)>0,\ell_{ng}\le d_{\text{walk}}} \ell_{ng}.
\end{equation*}

Departures and arrivals define transitions in the generator matrix $\mathcal{Q}(\tau)$ of the \gls{ctmc} associated to the demand process:
\begin{equation*}
\mathcal{Q}(\tau)[\mathbf{x},\mathbf{x}'] =
\begin{cases}
\lambda_{nm}(\tau), &  \text{if }\exists g_n(t),~\mathbf{x}'=\mathbf{x}+\Delta_{g_n(\mathbf{x}),m},\\
0, & \text{otherwise},
\end{cases}
\end{equation*}
where $\Delta_{g_n(\mathbf{x}),m}$ represents the departure of one active bike from the closest available node $g_n(\mathbf{x})$ and the creation of one in-transit bike toward $m$. Upon arrival, the number of traveling bikes decreases and the destination node receives either an active or depleted bike depending on $c_b(t)$. On the other hand, if $g_n(\mathbf{x})$ does not exist, i.e., there are no available bikes within walking distance, the user will be unable to reach their destination. These events are termed \emph{system availability failures}, or just failures for simplicity.

\subsection{Rebalancing model}
\label{sec:reb_model}

We consider a single rebalancing truck covering the whole area by traveling over $\mathcal{G}$. The truck's state at any given time is determined by its position $p(t)\in\mathcal{V}$ and current e-bike load $h(t) \in \{0,\dots,H_{\max}\}$, i.e., the number of e-bikes stored on the truck. We assume that the truck's battery is sufficient to cover the whole shift, as well as recharging any e-bikes that deplete their (far smaller) batteries.
At each node $n$, the truck can either load or unload bikes, where $\delta(t)\in\mathbb{Z}$ denotes the number of bikes deposited at time $t$, or recharge the e-bikes present at the node:

\begin{enumerate}
    \item[1)] Load or unload bikes:
    \begin{equation*}
        \begin{cases}
            o_n(t) \leftarrow o_n(t) + \delta(t),\\[2mm]
            h(t) \leftarrow h(t) - \delta(t),
        \end{cases}
    \end{equation*}
    with constraints $h(t)-H_{\max}\le\delta(t)\le h(t) $ and $\delta(t)\ge -o_n(t)$.
    
    \item[2)] Recharge bikes:
    each selected bike $b$ parked at node $n$ is recharged to full capacity, 
    \begin{equation*}
        c_b(t+\varepsilon) = C_{\max}^b.
    \end{equation*}
    In this case, we have
    \begin{equation*}
        o_n^c(t)\leftarrow o_n(t),\ o_n^d(t)\leftarrow 0.
    \end{equation*}
\end{enumerate}
The fleet size is also limited, as the total number of bikes in the system is limited by $O_{\max}$ to reduce operational costs.
\subsection{Fully dynamic rebalancing}
\label{sec:fully_dynamic_rebalancing}

State-of-the-art rebalancing schemes are either \emph{static}, with pre-determined rebalancing times, or \emph{dynamic}, with a rebalancing effort being triggered by specific conditions in the system. However, these methods entail a significant effort from multiple trucks, as the whole network needs to be rebalanced at once, and they lack the capability of correcting forecasting mistakes without triggering a full-scale rebalancing. The traveling time required by the trucks is also generally neglected.

In this work, we propose a \emph{fully dynamic} rebalancing scheme in which a single trucks continuously optimizes the system, performing local operations without the costs associated with a system-wide rebalancing and managing the evolution of the bike sharing system in real time. The instantaneous expected variation of bikes at node $n$ is then
\begin{equation*}
    \dot{S}_n(t) = \Lambda_n^{\text{in}}(\tau) - \Lambda_n^{\text{out}}(\tau)
    + u_n(t),
\end{equation*}
where $u_n(t)$ represents the contribution of rebalancing operations, being positive for drops and negative for pickups.
The rebalancing objective thus becomes to select the next truck action $a(t)\in\mathcal{A}$, where $\mathcal{A}$ denotes the set of feasible actions, that maintains service availability over all nodes while limiting travel and operational costs. In the following, this dynamic rebalancing task will be solved by a \gls{drl} framework capable of learning adaptive control policies directly from experience in the simulated environment.

\section{Proposed approach}
\label{sec:approach} 

The system is modeled as an infinite-horizon \gls{mdp} defined by
$\langle\mathcal{S},\mathcal{A},\rho,\mathbf{P},\gamma\rangle$, where $\mathcal{S}$ is the state space, $\mathcal{A}$ is the action space, $\rho:\mathcal{S}\times\mathcal{A}\times\mathcal{S}\to\mathbb{R}$ is the expected reward function, $\mathbf{P}\in[0,1]^{|\mathcal{S}|\times|\mathcal{A}|\times|\mathcal{S}|}$ is the action-dependent transition probability matrix, and $\gamma$ is the exponential discount factor.

\subsection{MDP formulation}

\begin{figure}[t!]
    \centering
    \includegraphics[width=0.35\textwidth]{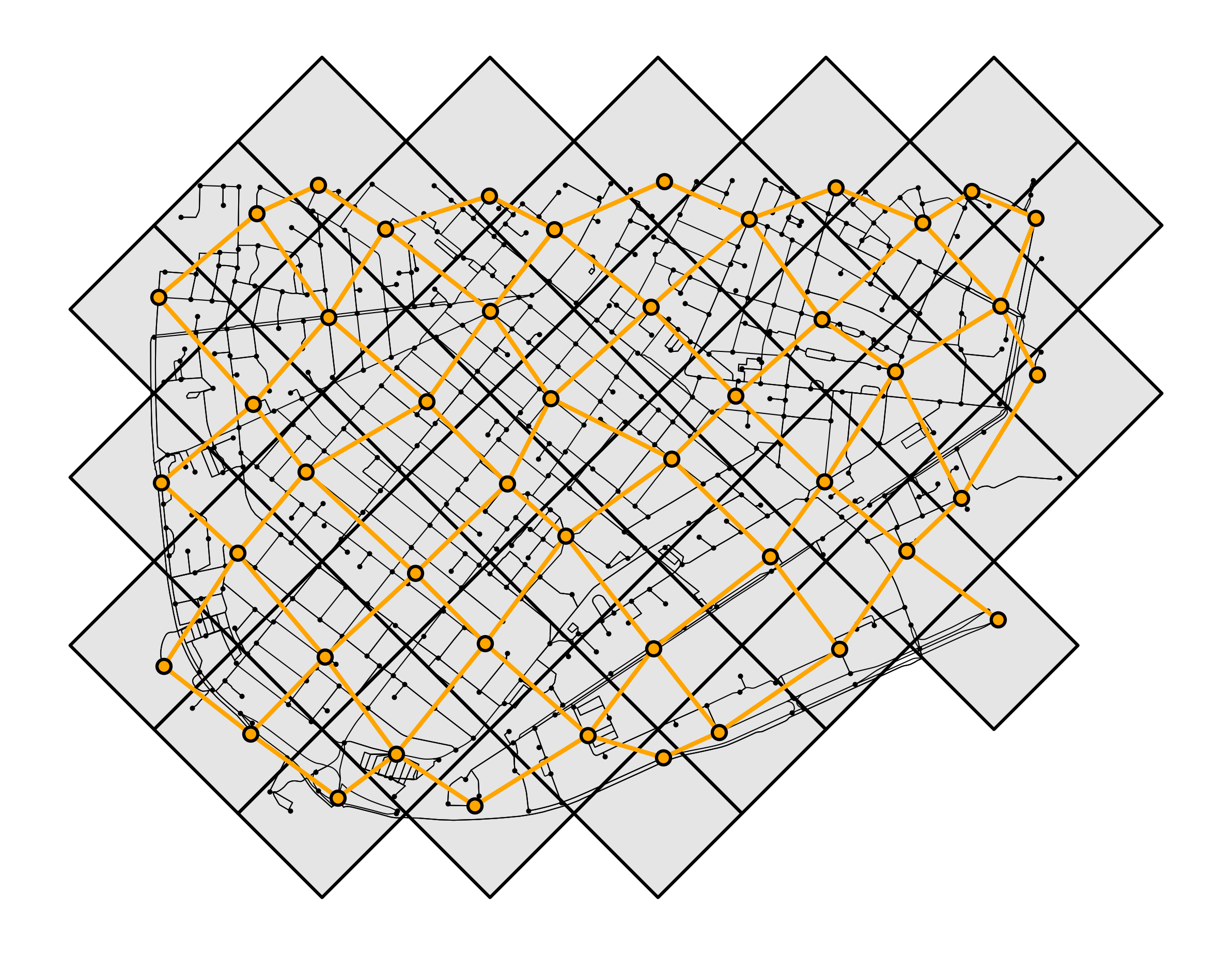}
    \caption{Road network and grid segmentation used for BSS simulation and truck navigation in Cambridge, MA.}
    \label{fig:graph}
\end{figure}

We assume the rebalancing truck to operate on a coarse-grained subgraph $\mathcal{G}'=(\mathcal{V}',\mathcal{E}')$ derived by aggregating $\mathcal{G}$ into a regular square grid. Each zone $v_i'$ is represented by a central node $z_i$, and adjacent zones are connected via weighted edges $w'(v_i',v_j')=\ell_{z_i z_j}$. The fine-scaled road network of the area we shall adopt in the simulations, with all street corners considered as possible bike locations,  and the coarse grid representation are shown in Fig.~\ref{fig:graph}. Additionally, the truck makes decisions over discrete time instants $k\in\mathbb{N}$, with a period $T_k$. This embedded \gls{dtmc} models the free evolution of the demand \gls{ctmc} in between subsequent actions.

The state $s(k)$ includes the truck load $h(kT_k)$, the previous action $a(k-1)$, the truck position in $\mathcal{G}'$, as well as a four-dimensional feature tuple for each node in $\mathcal{G}'$:
\begin{enumerate}
    \item[1.] A presence feature $\texttt{TP}$, which is $1$ if the truck is currently in that area and $0$ otherwise;
    \item[2.] The current number of charged bikes in the grid cell, $o^c_z(kT_k)=\sum_{n\in v_i'}o^c_n(kT_k)$;
    \item[3.] The time $\Delta k$ since the truck's last visit to the cell;
    \item[4.] A \emph{criticality score} $\psi(v_i',k)$, which represents a measure of the risk of failures happening in the cell.
\end{enumerate}
The criticality score is a proxy measure for indicating expected future shortages in the network, quantifying the rebalancing urgency for each area by predicting whether demand will exceed supply. The expected number of bikes in $v_i'$ in step $k'>k$ is
\begin{equation*} 
    \hat{o}_{v_i'}(k')=\sum\nolimits_{{n\in v_i'}} o_n^c(kT_k)+\sum\nolimits_{j=k}^{k'}T_k\left(\Lambda^{\text{\tiny in}}(\tau_j)-\Lambda^{\text{\tiny out}}(\tau_j)\right),
\end{equation*}
where $\tau_j$ is the time interval associated to step $j$.
The largest decrease in the number of available bikes over the next $K$ steps is then
\begin{equation*}
    d_{v_i'}^{\min}(k)=\sum\nolimits_{{n\in v_i'}}o_n^c(kT_k)-\min_{k'\in\{k+1,\ldots,k+K\}}\hat{o}_{v_i'}(k').
\end{equation*}
We then define parameter $\zeta_{v_i'}(k)$ as
\begin{equation*}
\zeta_{v_i'}(k)=\left(1+\alpha\left(\hat{o}_{v_i'}(k')-d_{v_i'}^{\min}(k)\right)\right)\left(1-\frac{o_{v_i'}(k)}{d_{v_i'}^{\min}(k)}\right),
\end{equation*}
where $\alpha$ is a tunable parameter. The criticality score is
\begin{equation*}
    \psi(v_i',k) =\exp(\zeta_{v_i'}(k))-1.
    \label{eq:critic_score_function}
\end{equation*}
Note that $|\psi(v_i',k)| \le 1\ \forall v_i',k$. A zone is classified as \textit{critical} if $\psi(v_i',k) > 0$, i.e., if $\zeta(v_i',k)<0$, which happens if the expected decrease in the number of bikes is larger than the current occupancy at any time. We also introduce a parameter $\sigma\in[0,1]$ distinguishing between \textit{stable} cells, in which $-\sigma \leq \psi(v_i',k) \leq 0$, and \textit{surplus} cells with $\psi(v_i',k) < -\sigma$, from which e-bikes can be safely removed. The truck state also includes the total criticality score $\Psi(k)$, defined as
\begin{equation*}
\Psi(k)=\sum_{v_i'\in\mathcal{G}'}I(\psi(v_i',k)>0)+(1-I(\psi(v_i',k)>0))\psi(v_i',k).
\end{equation*}
where $I(\cdot)$ is the indicator function, equal to $1$ if the argument is true and $0$ otherwise.

The actions available to the truck are then simple: it can deposit a fully charged e-bike, recharge or pick up a depleted e-bike in the current area, move to an adjacent cell over an edge in $\mathcal{E}'$, or take a void action, with the truck standing still and not performing any action.

The reward function is designed to limit failures while providing sufficient learning signals. Unweighted criticality scores are used so that rewards are independent of local demand, promoting \emph{spatial fairness} across cells. The reward $\rho(s,a_{\text{drop}},s')$ for dropping a bike in cell $v$ is
\begin{equation*}
\begin{aligned}
    \rho(s,a_{\text{drop}},s')=I(\psi(v,k)>0)(1+I(\psi(v,k+1)\le 0))\\
    +I(\psi(v,k)\le 0)(0.01-0.51 I(\psi(v,k)<-\sigma)).
\end{aligned}
\end{equation*}
The agent receives a reward of $1$ when dropping in a critical cell, increased to $2$ if the action restores stability, and a penalty of $-0.5$ when dropping in a surplus cell. Charging actions follow a similar logic, with an additional penalty for charging already high-battery bikes. The reward $\rho(s,a_{\text{pick-up}},s')$ for picking up a bike is
\begin{equation*}
\begin{aligned}
    \rho(s,a_{\text{pick-up}},s')=0.2 I(\psi(v,k)<-\sigma)-0.5I(\psi(v,k)>0)\\
    -2I(\psi(v,k)\le 0)I(\psi(v,k+1)>0).
\end{aligned}
\end{equation*}
This provides a small reward for removing bikes from surplus cells, a penalty for picking up from critical cells, and a stronger penalty when the action makes a stable cell critical. Movement actions are penalized when revisiting recently explored cells (based on $\Delta k$), unless unavoidable, while the null action is discouraged unless all cells are stable, i.e., $\psi(v_i',k)<0\ \forall v_i'\in\mathcal{V}'$. Additional penalties apply for invalid actions and empty-truck conditions. A global shaping term proportional to the variation in aggregate criticality encourages system-wide improvements, and a constant step penalty promotes efficiency.

Finally, actions may span multiple time steps due to travel and execution times. This is modeled via ``sticky’’ actions~\citep{sutton1999between}, which prevent the agent from acting again until completion.

\subsection{DRL algorithm and neural network architecture}
\label{subsec:ddqn}

We adopt the $n$-step return Double Deep Q-Network (DDQN) \citep{van2016deep} algorithm, through $\varepsilon$-greedy exploration with a Gaussian decay schedule:
\begin{equation*}
    \varepsilon(s) = \varepsilon_{\min} + (\varepsilon_{\max} - \varepsilon_{\min}) \, e^{-s^2 / \beta}
\end{equation*}
where $s$ denotes the cumulative number of training steps (time slots elapsed across all episodes), and $\beta$ is set so that $\varepsilon$ reaches $0.1$ after the first $\eta_{\varepsilon}$ fraction of total training steps.
A Graph Neural Network is employed for estimating the action-value function $Q_\theta$, since the state of the system can be naturally represented as a graph. More specifically, we exploit Graph Attention Networks (GATs)~\citep{velivckovic2017graph} to process spatially distributed zone features. Indeed, GATs extend Graph Convolutional Networks~\citep{kipf2016semi} by weighting neighboring zones according to learned attention coefficients, enabling the agent to focus on the most relevant regions.

\begin{figure}[t!]
    \centering
    \includegraphics[width=0.41\textwidth]{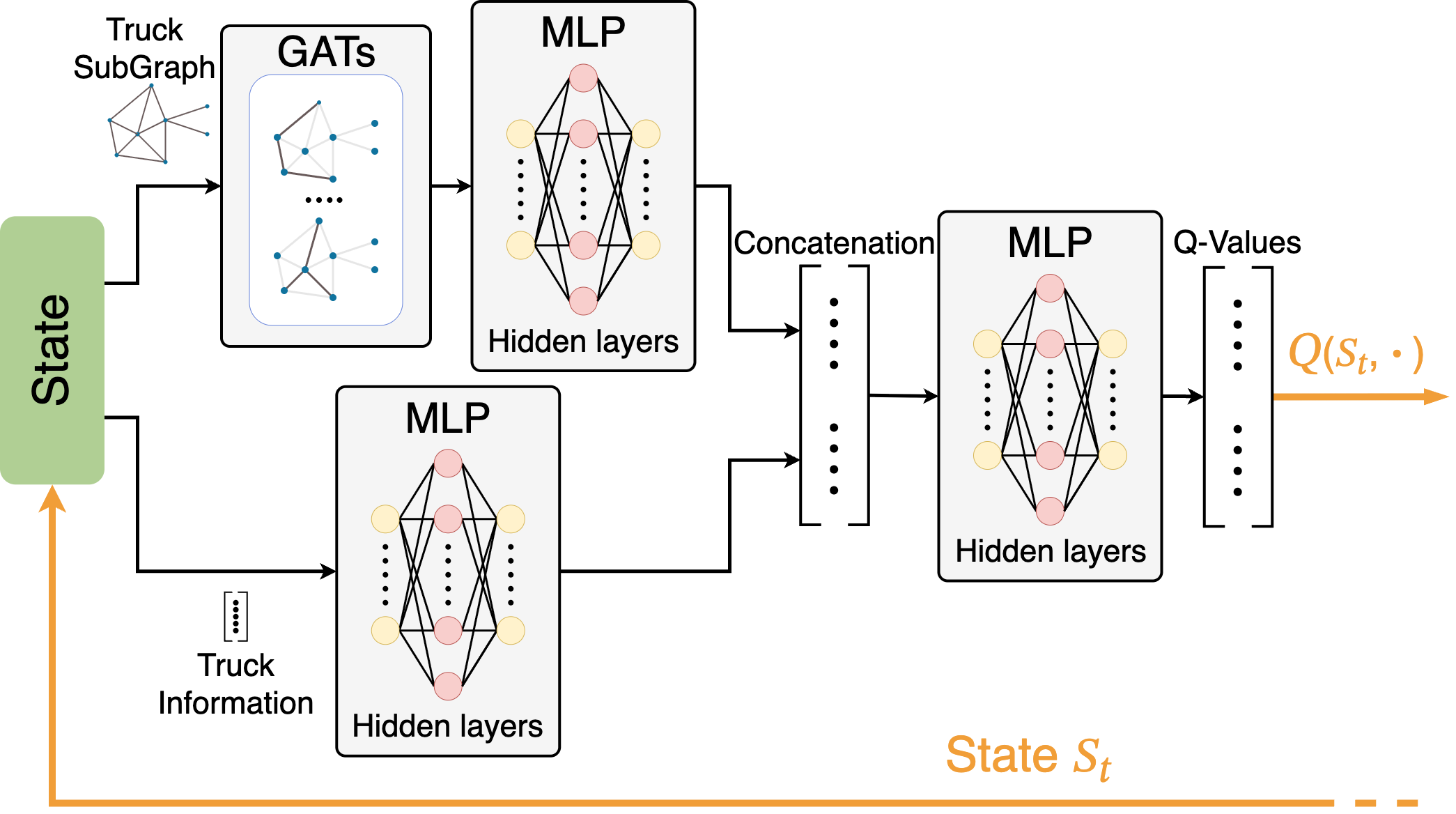}
    \caption{Neural network architecture of the DDQN agent.}
    \label{fig:dqn-module-architecture}
\end{figure}

The final architecture comprises three GAT layers with ReLU activations, followed by a global attention pooling layer that aggregates node-level features into a single graph embedding. This embedding is then processed by a multi-layer perceptron (MLP). A separate MLP encodes truck-specific features, and the two latent representations are concatenated and passed through a final MLP to produce the estimated $Q$-values for each action, as shown in Fig.~\ref{fig:dqn-module-architecture}. Lastly, as is typical in \gls{drl} scenarios, we leverage the Smooth $L_1$ loss function~\citep{girshick2015fast} and simple Stochastic Gradient Descent, which yields the most stable convergence in most \gls{drl} scenarios.

\section{Numerical results}\label{sec:results}
Our experimental analysis focuses on a representative area of approximately $4.6$~km$^2$ in Cambridge, MA (USA), which includes the \gls{mit}. This area corresponds to the realistic rebalancing area for a single truck, while containing several traffic hotspots.


The \gls{drl} algorithm was trained and evaluated over an \emph{ad hoc} discrete-event simulator.\footnote{For full reproducibility the source code is available at the following link: \url{https://github.com/edos08/BSS-DynamicRebalancing-RL}. All additional implementation details and non-explicit parameters are kept at their default values in the released code repository.} Both the \gls{drl} algorithm and the simulator are scalable to different or larger networks, albeit with potentially higher computational and training complexity. As benchmark, we use the well-known static rebalancing algorithm currently implemented by CitiBike in New York City~\citep{o2015data}, which we refer to as \gls{sr} in the following. No standard dynamic rebalancing baseline exists yet in the literature, making direct comparisons with other dynamic methods an open challenge.

\subsection{Dataset and experimental setup}


Trip data were sourced from the BlueBikes \gls{bss} operating in the Boston metropolitan area,\footnote{\url{https://bluebikes.com/system-data}} specifically from September and October 2022, corresponding to peak service usage months.
To capture temporal demand variations, data were segmented by day of week and divided into eight three-hour time slots ($\tau \in \{0,\dots,7\}$), with slot 0 beginning at 1 AM to better capture morning rush-hour dynamics. For each node pair $(n,m)$ and time slot $\tau$, departure rates were computed by summing all trips from $n$ to $m$ and normalizing by the number of days. Trips to and from outside the area were also modeled.

Since BlueBikes is a \textit{docked} system while our study addresses \textit{dockless} rebalancing, departure rates were interpolated to estimate potential demand across all graph nodes. Specifically, a $\vert\mathcal{V}\vert\times\vert\mathcal{V}\vert$ rate matrix was first filled on existing station entries, then completed with two-stage Inverse-Distance Weighting: missing destinations were estimated from nearby stations within 500 m, and inactive rows were inferred from neighboring active nodes. The matrix was finally rescaled to preserve the total departure rate. Notably, open large-scale dockless BSS datasets are scarce; docked systems like BlueBikes are the most viable alternative, and US cities’ relatively uniform demand patterns~\citep{weinreich2023bike} make IDW a sound approximation here.

Mean travel velocities and energy consumption estimates for the truck were extracted from the TomTom Boston Traffic Report.\footnote{\url{https://www.tomtom.com/traffic-index/boston-ma-traffic/}} The subgraph $\mathcal{G}'$ was built as a $300\times300$m square grid, as shown in Fig.\ref{fig:graph}. Episodes start on Monday at 1 AM, span 56 time slots, and initialize all bikes as fully charged, with at least $O_{\text{min}}=5$ bikes per cell in all runs. Key training hyperparameters are reported in Table~\ref{tab:hyperparams}.


\begin{table}[t]
\centering
\renewcommand{\arraystretch}{0.9}
\caption{Training hyperparameters.}
\begin{tabular}{|l|c||l|c|}
\hline
\textbf{Hyperparameter} & \textbf{Value} &
\textbf{Hyperparameter} & \textbf{Value} \\
\hline
Learning rate & $10^{-4}$ &
Batch size & 64 \\
Replay buffer size & $10^5$ &
Episodes (eps.) & 200 \\
Time slots per ep. & 56 &
Discount factor $\gamma$ & 0.95 \\
Initial $\varepsilon_{\max}$ & 1.0 &
Final $\varepsilon_{\min}$ & 0.01 \\
Explor. fraction $\eta_{\varepsilon}$ & 0.5 &
Soft update factor $\tau$ & 0.005 \\
\hline
\end{tabular}
\label{tab:hyperparams}
\end{table}

\subsection{Analysis of the results}

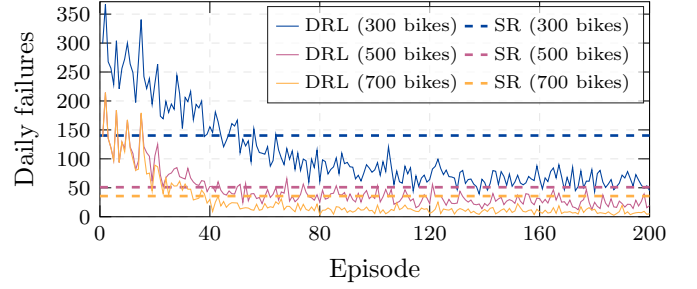
\begin{figure}[t]
\centering
\begin{tikzpicture}
\begin{axis}[
    width=1.0\columnwidth,
    height=0.5\columnwidth,
    xlabel={Episode},
    ylabel={Daily failures},
    xmin=-1, xmax=199,
    xtick={-1,39,79,119,159,199},
    xticklabels={0,40,80,120,160,200},
    ymin=0, ymax=2600,
    ytick={0,350,700,1050,1400,1750,2100,2450},
    yticklabels={0,50,100,150,200,250,300,350},
    ymajorgrids=true,
    xmajorgrids=true,
    grid style=dashed,
    legend style={legend columns=2,font=\scriptsize, anchor=north east, at={(0.99,0.98)}},
]

\addplot[
    color=color0
    ]
    table[x=episode,y=bike_300] {figures/training_comparison.dat};
\addlegendentry{DRL (300 bikes)};

\addplot[
    color=color0, line width=1pt, dashed
    ]
    table{
-1   982
199  982
};
\addlegendentry{SR (300 bikes)};

\addplot[
    color=color2
    ]
    table[x=episode,y=bike_500] {figures/training_comparison.dat};
\addlegendentry{DRL (500 bikes)};

\addplot[
    color=color2, line width=1pt, dashed
    ]
    table{
-1   357
199  357
};
\addlegendentry{SR (500 bikes)};

\addplot[
    color=color4
    ]
    table[x=episode,y=bike_700] {figures/training_comparison.dat};
\addlegendentry{DRL (700 bikes)};

\addplot[
    color=color4, line width=1pt, dashed
    ]
    table{
-1   250
199  250
};
\addlegendentry{SR (700 bikes)};

\end{axis}
\end{tikzpicture}
\caption{Average daily failures of \gls{drl} during training for different values of $O_{\max}$.}
\label{fig:convergence} 
\end{figure}

We now compare the \gls{drl} policy with the \gls{sr} baseline across different fleet sizes. In Fig.\ref{fig:convergence}, dashed lines denote the idealized \gls{sr} scheme, which redistributes bikes instantaneously at 1 AM and 1 PM following \cite{o2015data}, while solid lines denote the proposed \gls{drl} approach.

In the moderate scenario ($500$ bikes), \gls{drl} clearly surpasses \gls{sr} after convergence, reducing daily failures from $51$ to $12$. With $700$ bikes, our method reaches $3$ daily failures versus $35$ for \gls{sr}. Under scarcity ($300$ bikes), with fewer than 6 bikes per zone, \gls{drl} still improves performance, lowering failures from $140$ to $40$, though high failure rates persist. The performance gap favoring \gls{drl} widens as resources diminish, showing that fully dynamic rebalancing is most effective under limited flexibility due to real-time reaction to scarcity. Fleet size also affects learning speed: the agent converges in roughly $40$ episodes with $700$ bikes, while the $300$-bike case requires more than $100$ episodes and its performance is less stable after convergence, reflecting the difficulty of learning effective actions under tight constraints.

\begin{figure}[t]
\centering
\begin{tikzpicture}
\begin{axis}[
    width=1.0\columnwidth,
    height=0.45\columnwidth,
    xlabel={Time},
    ylabel={Hourly failures},
    xmin=0, xmax=55,
    ymin=0, ymax=54,
    ytick={0,12,24,36,48},
    yticklabels={0,4,8,12,16},
    xtick={2,5,10,13,18,21,26,29,34,37,42,45,50,53},
    xticklabels={{8.30AM,5.30PM,8.30AM,5.30PM,8.30AM,5.30PM,8.30AM,5.30PM,8.30AM,5.30PM,8.30AM,5.30PM,8.30AM,5.30PM}},
    x tick label style={rotate=45, anchor=east},
    ymajorgrids=true,
    xmajorgrids=true,
    grid style=dashed,
    legend style={legend columns=2, font=\scriptsize, anchor=north west, at={(0.36,0.98)}}
]

\addplot[color=lightgray, forget plot, name path=mon] table{
7.5 0
7.5 60
};

\addplot[color=lightgray, forget plot, name path=tue] table{
15.5 0
15.5 60
};
\addplot[color=lightgray, forget plot, name path=wed] table{
23.5 0
23.5 60
};
\addplot[color=lightgray, forget plot, name path=thu] table{
31.5 0
31.5 60
};
\addplot[color=lightgray, forget plot, name path=fri] table{
39.5 0
39.5 60
};
\addplot[color=lightgray, forget plot, name path=sat] table{
47.5 0
47.5 60
};

\addplot[
        fill=lightgray, 
        fill opacity=0.25,
        forget plot
    ]
    fill between[
        of=mon and tue
    ];

\addplot[
        fill=lightgray, 
        fill opacity=0.25,
        forget plot
    ]
    fill between[
        of=wed and thu
    ];

\addplot[
        fill=lightgray, 
        fill opacity=0.25,
        forget plot
    ]
    fill between[
        of=fri and sat
    ];

\addplot[
    color=color2, dashed, forget plot, name path=ben_min
    ]
    table[x=timeslot,y=bench_min] {figures/val_bench_evolution.dat};

\addplot[
    color=color2, dashed, forget plot, name path=ben_min
    ]
    table[x=timeslot,y=bench_min] {figures/val_bench_evolution.dat};

\addplot[
    color=color2, dashed, forget plot, name path=ben_max
    ]
    table[x=timeslot,y=bench_max] {figures/val_bench_evolution.dat};

\addplot[
        fill=color2, 
        fill opacity=0.25
    ]
    fill between[
        of=ben_min and ben_max
    ];
\addlegendentry{SR};

\addplot[
    color=color0, dashed, forget plot, name path=val_min
    ]
    table[x=timeslot,y=val_min] {figures/val_bench_evolution.dat};

\addplot[
    color=color0, dashed, forget plot, name path=val_max
    ]
    table[x=timeslot,y=val_max] {figures/val_bench_evolution.dat};

\addplot[
        fill=color0, 
        fill opacity=0.25
    ]
    fill between[
        of=val_min and val_max
    ];
\addlegendentry{DRL};

\addplot[
    color=color2
    ]
    table[x=timeslot,y=bench_mean] {figures/val_bench_evolution.dat};
\addlegendentry{SR (avg.)};

\addplot[
    color=color0
    ]
    table[x=timeslot,y=val_mean] {figures/val_bench_evolution.dat};
\addlegendentry{DRL (avg.)};

\end{axis}
\begin{axis}[
    width=1.0\columnwidth,
    height=0.45\columnwidth,
    ymin=0, ymax=34,
    xmin=0, xmax=55,
    xtick style={draw=none},
    ytick style={draw=none},
    yticklabels={draw=none},
    xtick={3.5, 11.5, 19.5, 27.5, 35.5, 43.5, 51.5},
    xticklabels={{Mon, Tue, Wed, Thu, Fri, Sat, Sun}},
    axis x line*=top]
\addplot[color=color1, forget plot] table{
0   60
55  60
};
    
\end{axis}

\end{tikzpicture}
\caption{Failures of \gls{drl} and \gls{sr} over time over $10$ episodes with $O_{\max}=500$.}
\label{fig:hourly} 
\end{figure}
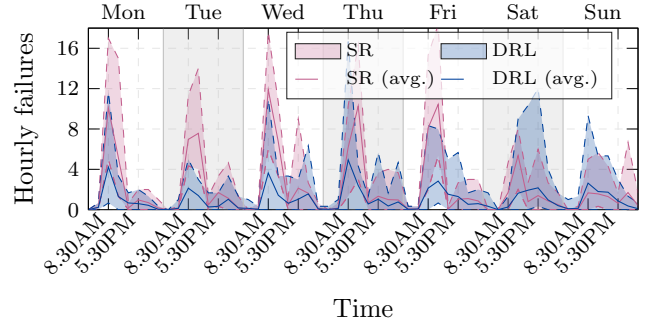

\begin{figure*}[t!] 
	\centering
	\subfigure[Normalized failures.]{\includegraphics[width=0.3\textwidth,trim={0cm 0cm 0cm 0cm},clip]{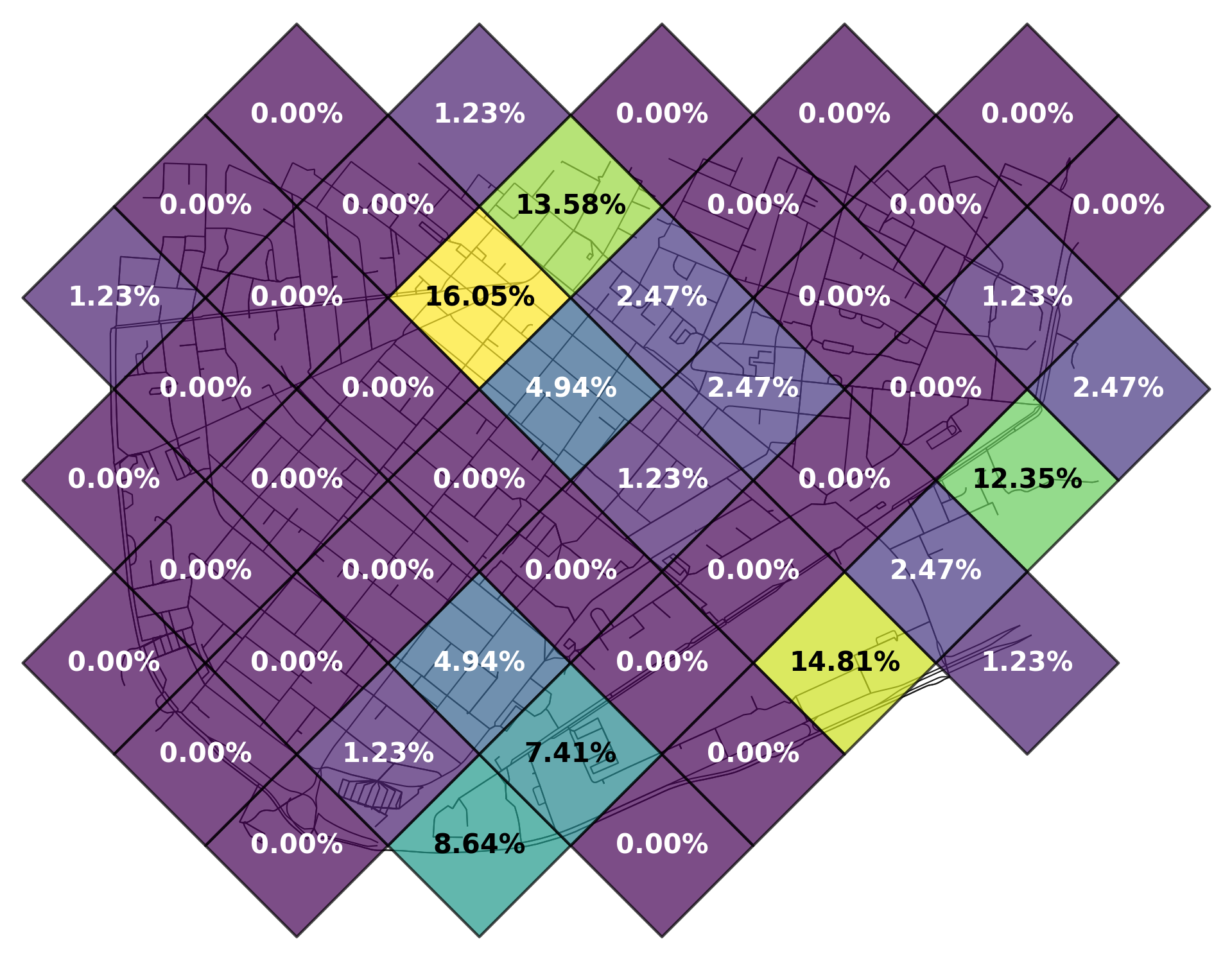}\label{fig:val-failures}} 
	\subfigure[Failure rates (failures/demand).]{\includegraphics[width=0.3\textwidth,trim={0cm 0cm 0cm 0cm},clip]{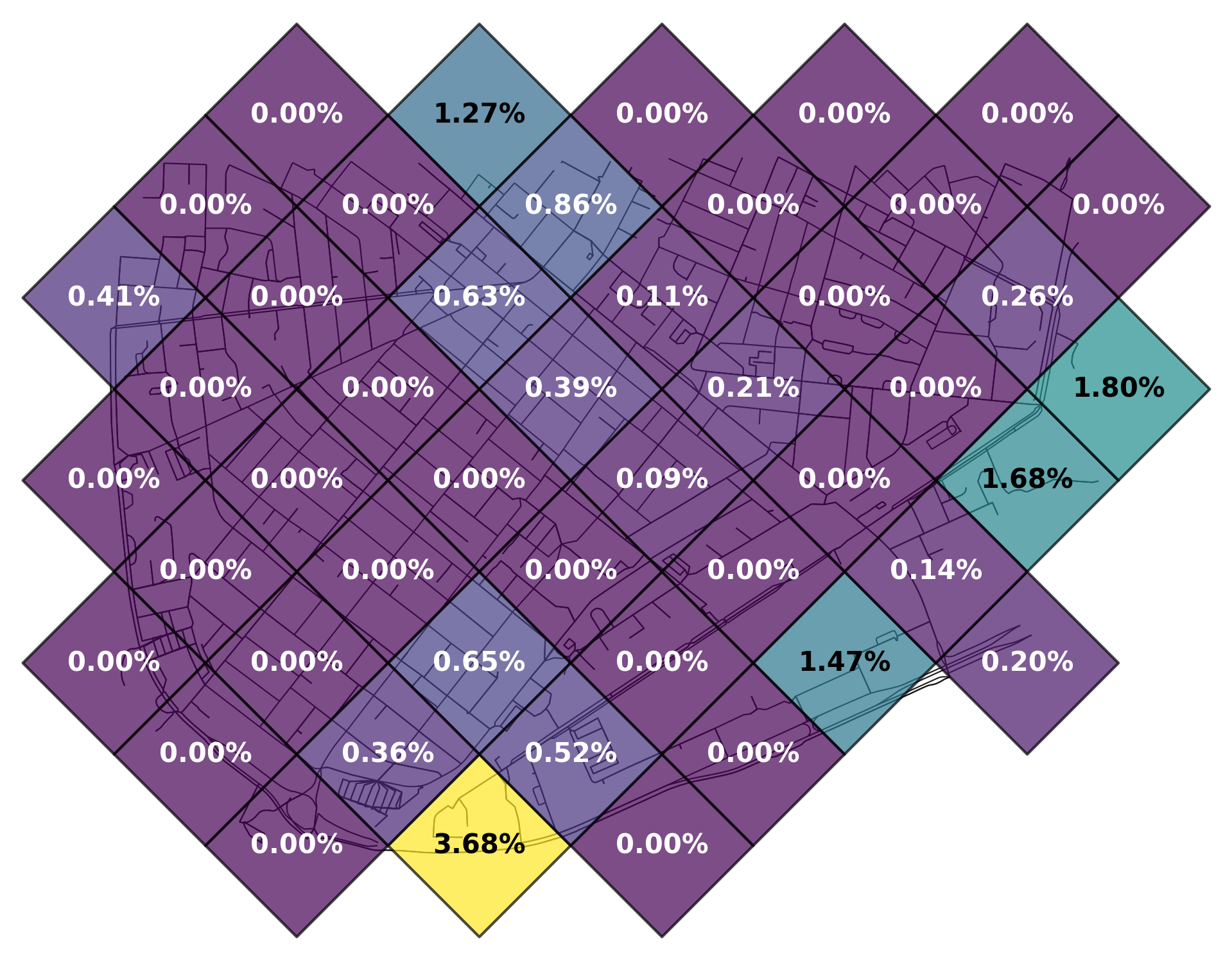}\label{fig:val-failure-rates}} 
	\subfigure[Normalized rebalancing actions.]{\includegraphics[width=0.3\textwidth,trim={0cm 0cm 0cm 0cm},clip]{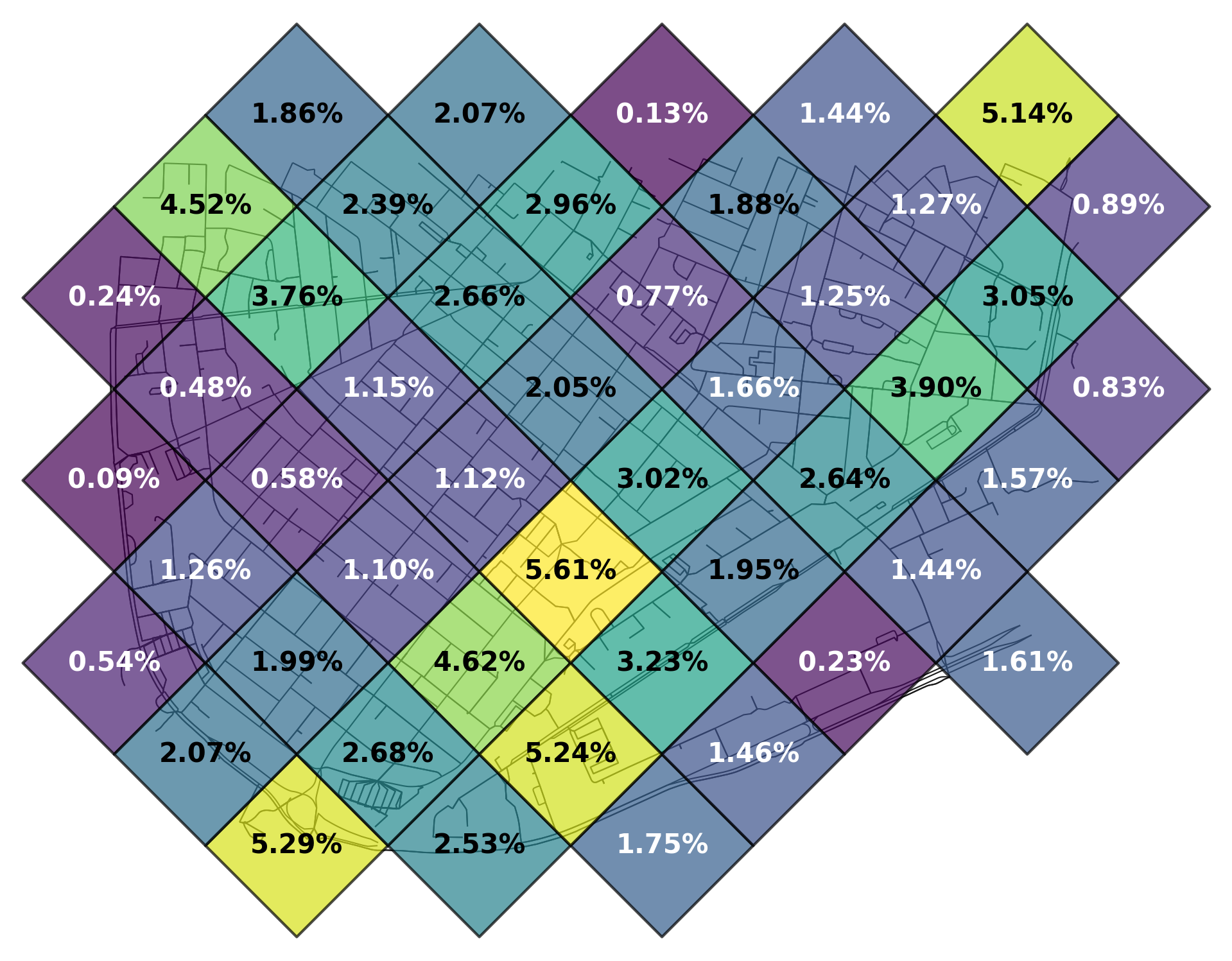}\label{fig:val-num-rebal-actions}} 
	\vspace{-.2cm}
	\caption{Spatial distribution of failures, failure rates, and rebalancing actions when using \gls{drl} with a fleet of $500$ bikes.}
	\label{fig:simulations}
\end{figure*}

Fig.~\ref{fig:hourly} shows the temporal evolution of failures over one episode with $O_{\text{max}} = 500$, using the best model after convergence. The \gls{drl} policy maintains consistently lower failure counts than \gls{sr} across the entire week. Results are averaged over 10 one-week episodes, and are reported both the mean trends and variability bands. As expected, clear rush-hour patterns emerge, with peaks around 8:30 AM and 5:30 PM for both methods. However, \gls{drl} systematically reduces both the magnitude and frequency of these peaks compared to \gls{sr}. While variability is still present—particularly during high-demand periods—\gls{drl} exhibits a more controlled behavior, with lower average failures and reduced extreme spikes, whereas \gls{sr} shows larger fluctuations and more pronounced peaks across multiple days.

A spatial breakdown of failures, failure rates, and rebalancing actions is reported in Fig.\ref{fig:simulations} for the $500$-bike configuration. Failures are highly localized: a small number of cells account for a significant fraction of total failures (peaking around $15\%$–$16\%$), scattered across high-demand zones rather than following a clear spatial corridor. When normalized by demand, failure rates remain near zero in most areas, with only a few cells exceeding $3\%$. This indicates that the system maintains relatively high service levels even in critical regions, and that failures result from localized demand peaks rather than widespread imbalance. Rebalancing actions are more diffuse, with the most serviced cells reaching $5\%$–$6\%$ of total actions. The agent allocates more effort to high-demand areas without exclusively targeting cells with the highest failures, indicating a proactive strategy that anticipates imbalances rather than only reacting to observed failures. The combination of sparse failures, low failure rates, and distributed rebalancing shows that the policy maintains global availability while selectively addressing localized fluctuations. Although high-demand cells naturally exhibit higher failure shares due to stochastic variability, no systematically neglected regions emerge, suggesting the fairness-oriented reward design mitigates low-demand area starvation and promotes balanced service\citep{cederle2025fairness}.

\section{Conclusions and outlook}\label{sec:conclusions}


This paper introduces a fully dynamic rebalancing framework for dockless bike sharing systems, combining a graph-based simulator with a \gls{drl} agent capable of acting in real time under stochastic demand. Using real-world data from Cambridge (MA), the proposed method significantly reduces service failures especially while operating with a limited fleet, and helps prevent localized shortages across the service area. These results highlight the potential of learning-based rebalancing to enhance availability while maintaining high operational efficiency.

Future work includes extending this framework to multi-truck settings, integrating user incentives to reduce operator workload, and exploring fairness-aware objectives to ensure equitable service across urban areas. 

\bibliography{ifacconf_REF}             
                                                   










\end{document}